\documentclass[english,pra]{revtex4}
\usepackage[T1]{fontenc}
\usepackage[latin1]{inputenc}
\usepackage{amssymb}

\makeatletter

\newcommand{\noun}[1]{\textsc{#1}}

\usepackage{amssymb}

\usepackage{babel}
\makeatother
\begin{document}

\title{Robustness of entangled states that are positive under partial transposition}

\author{Somshubhro Bandyopadhyay %
\footnote{Most of this work was completed while SB was a Research Associate
at the Institute for Quantum Information Science, University of Calgary.%
}}

\affiliation{DIRO, Universit\'{e} de Montr\'{e}al, C.P.6128, Succursale Centre-Ville,
Montr\'{e}al, Qu\'{e}bec H3C 3J7, Canada}

\email{bandyo@iro.umontreal.ca}

\author{Sibasish Ghosh}

\affiliation{The Institute of Mathematical Sciences, C. I. T Campus, Taramani,
Chennai 600113, India}

\email{sibasish@imsc.res.in}

\author{Vwani Roychowdhury}

\affiliation{Electrical Engineering Department, UCLA, Los Angeles, CA 90095, USA}

\email{vwani@ee.ucla.edu}

\begin{abstract}
We study robustness of bipartite entangled states that are positive
under partial transposition (PPT). It is shown that almost all PPT
entangled states are unconditionally robust, in the sense, both inseparability
and positivity are preserved under sufficiently small perturbations
in its immediate neighborhood. Such unconditionally robust PPT entangled
states lie inside an open PPT entangled ball. We construct examples
of such balls whose radii are shown to be finite and can be explicitly
calculated. This provides a lower bound on the volume of all PPT entangled
states. Multipartite generalization of our constructions are also
outlined. 
\end{abstract}
\maketitle

\section{Introduction}

Robustness of an entangled quantum state quantifies its ability to
remain inseparable/entangled in the presence of decoherence-- that
is, how much noise can be added before the entangled state becomes
separable \cite{Vidal99}-\cite{Caval-Cunha2006}. Recently it was
shown that weakly entangled states are dense and robust, and in particular,
bound entangled states constructed from an unextendible product basis
(UPB) \cite{bennett99} are \emph{conditionally} robust, in the sense
that sufficiently small perturbations along certain directions preserve
both inseparability and the positivity under partial transposition
(PPT) properties \cite{orus04}. While this is a significant result,
robustness of generic bound entangled states \cite{boundent} (bound
entangled states are assumed to be PPT unless otherwise stated) ,
i.e., preservation of their (a) inseparability and (b) positivity
under partial transposition, in their immediate neighborhood under
sufficiently small perturbation, is not well understood.

Consider a bipartite quantum system $AB$, described by the joint
Hilbert space ${\mathcal{H}}={\mathcal{H}}_{A}\otimes{\mathcal{H}}_{B}$,
an inseparable PPT density matrix $\rho\in\mathcal{H}$, an arbitrary
perturbation of $\rho$: \begin{equation}
\rho^{\prime}=\frac{1}{1+\epsilon}(\rho+\epsilon\sigma)\label{robust}\end{equation}

where $\sigma$ is any other density matrix and $\epsilon>0$ is an
infinitesimal noise parameter. We say that $\rho$ is unconditionally
robust if and only if it is always inside a PPT ball, that is, for
any sufficiently small perturbation along an arbitrary direction the
state remains PPT, and inseparable.

The question, whether a given bound entangled state is unconditional
robust, is a non-trivial one. If we choose $\sigma$ in the above
equation to be a PPT state, then although PPT property is surely preserved
for any choice of $\epsilon$, it doesn't guarantee that the perturbed
state remains inseparable. On the other hand, if $\sigma$ is chosen
to be an entangled state with a non-positive spectrum under partial
transposition (NPT), then it is possible that the perturbed state
becomes distillable for any choice of $\epsilon$., thereby losing
the PPT property. In fact such examples have been found, although
in a different context \cite{Vidal2002}.

We prove that any PPT entangled state is either inside or on the surface
of a closed PPT entangled ball. Thus, almost all PPT entangled states
are unconditionally robust, and those on the surface of such balls
are conditionally robust. The radius of such a PPT entangled ball
may be suitably defined using an appropriate distance measure (trace
norm or Bures norm or Hilbert-Schmidt norm) between the centre-of-the-ball-state
and the states that are on the surface of the ball. A corollary of
the above result is that almost all PPT states are unconditionally
robust.

We provide examples where the radius of PPT entangled balls, constructed
in the neighbourhood of bound entangled states from an unextendible
product basis \cite{bennett99} (such bound entangled states are denoted
by BE-UPB), are shown to be finite and can be explicitly calculated.

Moreover, we show that bound entangled states can also be maximally
robust in certain directions. That is, one can mix a bound entangled
state with certain product states, such that the mixture remains bound
entangled as long as the proportion of the bound entangled state is
non-zero.

Finally, we prove that for every BE-UPB state (i.e., an edge BE state
\cite{lewenstein2001}), there is a region such that a mixture (the
coefficients of such a mixture is bounded) of an BE-UPB state with
any separable state is bound entangled inside the region\emph{.} This
may be considered as dual to the result---every PPT entangled state
can be expressed as a mixture of a separable state with an edge PPT
entangled state---obtained in Ref. \cite{lewenstein2001}.

\section{background}

Consider a bipartite quantum system $AB$, described by the joint
Hilbert space ${\mathcal{H}}={\mathcal{H}}_{A}\otimes{\mathcal{H}}_{B}$,
where dimensions of ${\mathcal{H}}_{A},\mathcal{H}_{B}$ are $d_{1},d_{2}$
respectively. Let ${\mathcal{D}}$ be the set of density matrices
of the system $AB$, and ${\mathcal{B}}$ be the set of linear operators
on ${\mathcal{H}}$. Thus ${\mathcal{D}}$ is a convex subset of the
$\left(d_{1}d_{2}\right)^{2}$-dimensional space ${\mathcal{B}}$.
Let ${\mathcal{S}}$ be the set of all separable states. Thus ${\mathcal{S}}$
is a convex as well as compact (with respect to usual metrics like
trace norm, or Hilbert-Schmidt norm, etc.) subset of ${\mathcal{D}}$.

Let $\left\{ |i\rangle_{A}:i=1,2,\ldots,d_{1}\right\} $, $\left\{ |j\rangle_{B}:j=1,2,\ldots,d_{2}\right\} $
be the standard orthonormal basis of ${\mathcal{H}}_{A}$, ${\mathcal{H}}_{B}$
respectively. The partial transpose ${\rho}^{T_{B}}$ of any $\rho\in{\mathcal{D}}$
(defined with respect to the standard orthonormal product basis $\left\{ |i\rangle_{A}\otimes|j\rangle_{B}:i=1,2,\ldots,d_{1};j=1,2,\ldots,d_{2}\right\} $
of ${\mathcal{H}}$), is given by\begin{equation}
_{B}\langle j|\otimes_{B}\langle i|{\rho}^{T_{B}}|i^{\prime}\rangle_{A}\otimes|j^{\prime}\rangle_{B}\equiv~_{B}\!{\langle}j^{\prime}|\otimes_{B}\langle i|{\rho}|i^{\prime}\rangle_{A}\otimes|j\rangle_{B}\label{eq:}\end{equation}
 for all $i,i^{\prime}\in\{1,2,\ldots,d_{1}\}$ and for all $j,j^{\prime}\in\{1,2,\ldots,d_{2}\}$.
Let ${\mathcal{P}}$ be the set of all elements $\rho$ of ${\mathcal{D}}$,
such that ${\rho}^{T_{B}}\geq0$. Thus ${\mathcal{S}}$ is a proper
subset of ${\mathcal{P}}$ whenever $d_{1}d_{2}\geq8$.

Throughout this paper we will extensively use the theory of entanglement
witness. Here we provide a brief review of the pertinent results.
We begin with the definition of entanglement witness \cite{Horodecki96,terhal00,lewenstein2000}
and discuss some of its properties.

\textbf{Definition 1 (Entanglement Witness)} \textit{\emph{An entanglement
witness $W$ is a member of ${\mathcal{B}}$ such that}}

(i) \textit{\emph{$W=W^{\dagger}$}}\emph{, }

(ii) \textit{\emph{${\textrm{Tr}}(W\sigma)\geq0$ for all $\sigma\in{\mathcal{S}}$}}\emph{, }

(iii) \textit{\emph{there exists at least one entangled state $\rho$
of $AB$ such that ${\textrm{Tr}}(W\rho)<0$}}\emph{,} \textit{\emph{and}}

(iv) \textit{\emph{${\textrm{Tr}}(W)=1$}} \cite{note1}.

If $W$ is an entanglement witness and $\rho$ is an entangled state
such that ${\textrm{Tr}}(W\rho)<0$, then we say $W$ \textit{witnesses}
(or \textit{detects}) the entanglement in $\rho$. For each entanglement
witness $W$, one can write the spectral decomposition as:\begin{equation}
W=\sum_{i=1}^{p}{\lambda}_{i}^{+}|e_{i}^{+}\rangle\langle e_{i}^{+}|-\sum_{j=1}^{n}{\lambda}_{j}^{-}|e_{j}^{-}\rangle\langle e_{j}^{-}|,\label{witness-decomp}\end{equation}
 where ${\lambda}_{i}^{+}$'s are positive eigenvalues of $W$ with
corresponding eigenvectors $|e_{i}^{+}\rangle$ for $i=1,2,\ldots,p$
($p$ a positive integer) and $-{\lambda}_{j}^{-}$'s are negative
eigenvalues of $W$ with corresponding eigenvectors $|e_{j}^{-}\rangle$
for $j=1,2,\ldots,n$ ($n$ a positive integer). Thus $W=W^{+}-W^{-}$
and\begin{eqnarray}
\textrm{Tr}(W) & = & {\textrm{Tr}}(W^{+})-~{\textrm{Tr}}(W^{-})\nonumber \\
 & = & \sum_{i=1}^{p}{\lambda}_{i}^{+}-\sum_{j=1}^{n}{\lambda}_{j}^{-}=1.\label{Tr(W)}\end{eqnarray}
 For all density matrices $\pi\in{\mathcal{D}}$,\begin{equation}
-{\textrm{Tr}}(W^{-})\leq{\textrm{Tr}}(W\pi)\leq{\textrm{Tr}}(W^{+}).\label{Tr(W)bound}\end{equation}
 $W^{+}$ is therefore called the \textit{positive} part of $W$ and
$W^{-}$ is called the \textit{negative} part of $W$. Note that both
$p,n\geq1$ and the $n$ dimensional subspace spanned by the eigenvectors
$|e_{j}^{-}\rangle$ for $j=1,2,\ldots,n$, contains no product state.

\textbf{Lemma 1} \textit{\emph{\cite{Horodecki96,lewenstein2001}
Let $\rho$ be any given entangled state in ${\mathcal{H}}_{A}\otimes{\mathcal{H}}_{B}$,
where dim ${\mathcal{H}}_{A}=d_{1}$, dim ${\mathcal{H}}_{B}=d_{2}$,
and $d_{1}d_{2}\geq8$. There exists an entanglement witness $W_{\rho}$
such that}}

(i) \textit{\emph{${\textrm{Tr}}(W_{\rho}\rho)<0$, and}}

(ii) \textit{\emph{there also exists a separable state ${\sigma_{\rho}}$
such that ${\textrm{Tr}}\left(W_{\rho}{\sigma}_{\rho}\right)=0$.}}

Let $\rho$ be any state of $AB$, taken from $({\mathcal{P}}-{\mathcal{S}})$.
Let ${\mathcal{W}}_{\rho}$ be the collection of all entanglement
witnesses such that $\textrm{Tr}(W\rho)<0,W\in\mathcal{W}_{\rho}$.
${\mathcal{W}}_{\rho}$ is a non-empty subset of ${\mathcal{B}}$
as for each entangled state $\rho\in\mathcal{D}$ there exists at
least one entanglement witness \cite{guhne}. For each $W\in{\mathcal{W}}_{\rho}$,
let $\mathcal{D}_{\textrm{w}}$ be the set of all entangled density
matrices of $AB$, whose inseparability is witnessed by $W$. For
two entanglement witnesses $W_{1},W_{2}\in\mathcal{W}$,$W_{2}$ is
said to be \textit{finer} than $W_{1}$ if $D_{\textrm{w}_{1}}$ is
a subset of $D_{\textrm{w}_{2}}$. An element $W_{\rho}\in{\mathcal{W}}_{\rho}$
is an \textit{optimal} entanglement witness \cite{lewenstein2001}
for $\rho$ if there is no $W\in{\mathcal{W}}_{\rho}$ which is finer
than $W_{\rho}$.

\textbf{Definition 2} \textbf{\textit{\emph{(Edge state)}}} \textit{\emph{An
element ${\delta}\in({\mathcal{P}}-{\mathcal{S}})$ is said to be
an edge state if there is no product state $|\psi\rangle_{A}\langle\psi|\otimes|\phi\rangle_{B}\langle\phi|\in{\mathcal{S}}$
and there is a positive number $\epsilon$ such that $\delta-\epsilon|\psi\rangle_{A}\langle\psi|\otimes|\phi\rangle_{B}\langle\phi|$
is a positive operator on ${\mathcal{H}}_{A}\otimes{\mathcal{H}}_{B}$
or is positive under partial transposition or both\cite{lewenstein2000,lewenstein2001}.}}

It was shown in \cite{lewenstein2001} that for any $\rho\in({\mathcal{P}}-{\mathcal{S}})$,
there exist an element $\sigma\in{\mathcal{S}}$, an edge state $\delta\in({\mathcal{P}}-{\mathcal{S}})$,
and a number $\Lambda\in[0,1]$ such that $\rho=\Lambda\sigma+(1-\Lambda)\delta$,
and for fixed $\delta$, this representation is optimal (in the sense
that one cannot increase $\Lambda$ by subtracting a non-zero factor
of the projector of a product state from $\delta$). Thus by choosing
the {}``nearest'' \cite{note2} separable state ${\sigma}_{\rho}$
(of $\rho$), one can expect to select an edge state ${\delta}_{\rho}\in({\mathcal{P}}-{\mathcal{S}})$
such that $\rho={\Lambda}_{\rho}{\sigma}_{\rho}+\left(1-{\Lambda}_{\rho}\right){\delta}_{\rho}$,
where ${\Lambda}_{\rho}$ is the largest achievable value. Note that
the BE-UPB states \cite{bennett99} are the edge states.

\section{Results}

This section is arranged as follows: We first introduce the necessary
definitions and then prove the results on unconditional robustness
of PPT entangled states

\textbf{Definition 3 (Non-empty ball around a density matrix)}

\textit{\emph{For any $\rho\in{\mathcal{D}}$ and any $\lambda\in(0,1]$,
a non-empty ball $B(\rho;\lambda)$ of radius $\lambda$ around $\rho$
is defined as $B(\rho;\lambda)=\left\{ \mu{\rho}^{\prime}+(1-\mu)\rho:{\rho}^{\prime}\in{\mathcal{D}}~{\textrm{and}}~0\leq\mu<\lambda\right\} $. }}

\textbf{Definition 4 (neighbourhood robustness)}

\textit{\emph{A PPT entangled state $\rho\in{\mathcal{D}}$ is}}

\textit{\emph{(i)}} \textit{maximally robust} \textit{\emph{if there
exists a member ${\sigma}\in{\mathcal{D}}$ such that $x\sigma+(1-x)\rho$
is a PPT entangled state for all $x\in[0,1[$.}}

\textit{\emph{(ii)}} \textit{robust relative \cite{Vidal99} to} \textit{\emph{$T$
if there exist a non-empty subset $T$ of ${\mathcal{D}}$ and an
element $z_{0}\in\;]0,1[$ such that the states $z\sigma+(1-z)\rho$
are PPT bound entangled for all $\sigma\in T$ and for all $z\in[0,z_{0}[$}}\emph{,}

\textit{\emph{(iii)}} \textit{unconditionally robust} \textit{\emph{if
there exists a non-empty ball $B(\rho;\lambda)$ containing only PPT
entangled states}}\emph{, }

\textbf{Lemma 2} $B(I/D;1/(D-1))$ is a separable ball.

\textbf{Proof}. In Ref. \cite{gurvits02} it was shown that $\rho\in{\mathcal{D}}$
is separable if its purity, \textit{i.e.}, ${\textrm{Tr}}\left({\rho}^{2}\right)$,
is less than $\frac{1}{D-1}$, where $D=d_{1}d_{2}$. Applying this
to an arbitrary element ${\rho}_{\mu}\equiv\mu{\rho}^{\prime}+(1-\mu)\frac{I}{D}$
of $B(I/D;\lambda)$, it follows that ${\rho}_{\mu}$ is separable
if\begin{equation}
~{\textrm{Tr}}({\rho}_{\mu}^{2})=\frac{1}{D}+{\mu}^{2}\left(~{\textrm{Tr}}\left({{\rho}^{\prime}}^{2}\right)-\frac{1}{D}\right)<\frac{1}{D-1}\label{GB-bound}\end{equation}
 for all elements ${\rho}^{\prime}$ of ${\mathcal{D}}$. Thus $\frac{1}{D}+{\mu}^{2}\left(1-\frac{1}{D}\right)$
must be less than $\frac{1}{D-1}$, \textit{i.e.}, $\mu<\frac{1}{D-1}$.
Hence, every element of the ball $B(I/D;1/(D-1))$ is separable \cite{note3}.
$\square$

\textbf{Definition 5 (cut cone)} Let $\rho\in({\mathcal{P}}-{\mathcal{S}})$.
Consider the cone $K_{S}\equiv\{\mu\sigma+(1-\mu)\rho:0\leq\mu\leq1~{\textrm{and}}~\sigma\in{\mathcal{S}}\}$.
Let $\lambda\in\;]0,1]$. Then the set $K_{S}\bigcap B(\rho;\lambda)$
is called the cut cone of height $\lambda$, with vertex at $\rho$
and is denoted by $K_{S}(\rho;\lambda)$.

\subsection{Construction of a class of PPT entangled states }

Let $\rho$ be any given element of $({\mathcal{P}}-{\mathcal{S}})$.
Then from part (i) of Lemma 1, there exists an entanglement witness
$W_{\rho}$ such that ${\textrm{Tr}}\left(W_{\rho}\rho\right)$ $=-{\lambda}_{\rho}$,
Define the following family of states: \begin{equation}
{\mathcal{F}}_{\rho}=\left\{ {\rho}_{x}\in{\mathcal{D}}:{\rho}_{x}=x\rho+(1-x)\frac{\textrm{I}}{\textrm{D}}~{\textrm{and}}~0\leq x\leq1\right\} ,\label{frho}\end{equation}
 subset of ${\mathcal{P}}$. Now,\begin{equation}
{\textrm{Tr}}\left(W_{\rho}{\rho}_{x}\right)=\frac{1}{D}-x\left(\frac{1}{D}+{\lambda}_{\rho}\right)<0,\,\forall x\in\;]\frac{1}{(1+D\lambda_{\rho})},1].\label{nexteq1}\end{equation}
 Thus,\begin{equation}
{\rho}_{x}\in({\mathcal{P}}-{\mathcal{S}})\,\forall x\in\;]\frac{1}{1+D\lambda_{\rho}},1].\label{rhox}\end{equation}
 Consider the following subfamily of ${\mathcal{F}}_{\rho}$:\begin{equation}
{\mathcal{F}}_{\rho}^{1/(1+D{\lambda}_{\rho})}=\left\{ {\rho}_{x}\in{\mathcal{F}}_{\rho}:\frac{1}{1+D{\lambda}_{\rho}}<x\leq1\right\} .\label{frhod}\end{equation}
 Thus all elements of ${\mathcal{F}}_{\rho}^{1/(1+D{\lambda}_{\rho})}$
are PPT entangled states.

\subsection{Unconditional Robustness}

We now select an arbitrary element $\rho_{x}\in{\mathcal{F}}_{\rho}^{1/(1+D{\lambda}_{\rho})}$
and construct the following family of density matrices\begin{equation}
{\mathcal{G}}_{{\rho},1/(1+D{\lambda}_{\rho})}=\left\{ \tau(\rho,\sigma,x,y)\in{\mathcal{D}}:\tau(\rho,\sigma,x,y)=y{\sigma}+(1-y){\rho}_{x},~{\textrm{for}}~{\rho}_{x}\in{\mathcal{F}}_{\rho}^{1/(1+D{\lambda}_{\rho})},~\sigma\in{\mathcal{D}},~y\in[0,1)\right\} .\label{grhod}\end{equation}
 For any $\tau(\rho,\sigma,x,y)\in{\mathcal{G}}_{{\rho},1/(1+D{\lambda}_{\rho})}$,
we have\begin{equation}
\tau(\rho,\sigma,x,y)=(1-s(x,y))\left\{ t(x,y)\sigma+(1-t(x,y))\frac{I}{D}\right\} +s(x,y)\rho,\label{eqn1}\end{equation}
 where\begin{eqnarray}
s(x,y) & = & 1-x(1-y)\label{s(xy)}\\
t(x,y) & = & y/s(x,y)\label{t(xy)}\end{eqnarray}
 for $(x,y)\in\textrm{\,}]1/(1+D{\lambda}_{\rho}),1]\times[0,1[$.

The function $t(x,y)$ is well-defined only for $(x,y)\in\textrm{\,}]1/(1+D{\lambda}_{\rho}),1[\times[0,1[$,
and in that case, the range of $t(x,y)$ is $[0,1[$. Also the range
of $s(x,y)$ is $]0,1[$ whenever $(x,y)\in\textrm{\,}]1/(1+D{\lambda}_{\rho}),1[\times[0,1[$.
From a result in \cite{gurvits02}, it follows that the density matrix
$t(x,y)\sigma+(1-t(x,y))\frac{I}{D}$ is separable for all $\sigma\in{\mathcal{D}}$
provided $t(x,y)<1/(D-1)$, \textit{i.e.}, $y\in[0,(1-x)/(D-1-x)[$
whenever $x\in]1/(1+D{\lambda}_{\rho}),1[$. Thus $\tau(\rho,\sigma,x,y)$,
given in eqn.$\,$(\ref{eqn1}), is PPT for all $\sigma\in{\mathcal{D}}$
such that $y\in[0,(1-x)/(D-1-x)[$ whenever $x\in\,]1/(1+D{\lambda}_{\rho}),1[$.
Now \[
{\textrm{Tr}}\left(W_{\rho}\tau(\rho,\sigma,x,y)\right)=y\left\{ {\textrm{Tr}}\left(W_{\rho}\sigma\right)+\frac{x\left(1+D{\lambda}_{\rho}\right)-1}{D}\right\} -\frac{x\left(1+D{\lambda}_{\rho}\right)-1}{D}\]
\begin{equation}
=y{\textrm{Tr}}\left(W_{\rho}\sigma\right)-(1-y)\frac{x\left(1+D{\lambda}_{\rho}\right)-1}{D}.\label{eqn2}\end{equation}
 We have \begin{equation}
{\textrm{Tr}}\left(W_{\rho}\sigma\right)+\frac{x\left(1+D{\lambda}_{\rho}\right)-1}{D}\leq\frac{1}{p(W_{\rho})}~{\textrm{Tr}}\left(W_{\rho}^{+}\right)+\frac{x\left(1+D{\lambda}_{\rho}\right)-1}{D},\label{nexteq2}\end{equation}
 for all $\sigma\in{\mathcal{D}}$ where $W_{\rho}^{+}$ is the positive
part of $W_{\rho}$. Thus, for all $\sigma\in{\mathcal{D}}$, $\tau(\rho,\sigma,x,y)$
is a PPT entangled state provided $x\in\,]1/(1+D{\lambda}_{\rho}),1[$
and $y\in[0,y_{0}(x)[$ where \begin{equation}
y_{0}(x)=~{\textrm{min}}\left\{ \frac{1-x}{D-1-x},~\frac{p(W_{\rho})\left\{ x\left(1+D{\lambda}_{\rho}\right)-1\right\} }{D~{\textrm{Tr}}\left(W_{\rho}^{+}\right)+p(W_{\rho})\left\{ x\left(1+D{\lambda}_{\rho}\right)-1\right\} }\right\} .\label{eqn3}\end{equation}
 We can therefore state,

\textbf{Theorem 1 (unconditional robustness)} \textit{\emph{For any
PPT bound entangled state $\rho$ and for each $x\in]1/(1+D{\lambda}_{\rho}),1[$,
the ball $B({\rho}_{x};y_{0}(x))$ contains only PPT entangled states,
where $y_{0}(x)$ is given in equation (\ref{eqn3}); ${\lambda}_{\rho}$
is a positive number where ${\textrm{Tr}}\left(W_{\rho}\rho\right)=-{\lambda}_{\rho}$.}}

\textbf{Remark 2} Each member of the set \begin{equation}
{\tilde{\mathcal{F}}}_{\rho}^{1/(1+D{\lambda}_{\rho})}=\left({\mathcal{F}}_{\rho}^{1/(1+D{\lambda}_{\rho})}-\{\rho\}\right)\label{deleted}\end{equation}
 is an unconditionally robust PPT entangled state. Also $y_{0}(x)$
(given in eqn.$\,$(\ref{eqn3})) provides a lower bound on the maximum
size of the ball (containing only PPT bound entangled states) around
${\rho}_{x}$ for each $x\in\,]1/(1+D{\lambda}_{\rho}),1[$. The largest
range of $x$ can be obtained by taking the maximum possible value
of ${\lambda}_{\rho}$ (for example, as given in Lemma 1). However
$x$ cannot be arbitrarily close to $0$, as for all such $x$, ${\rho}_{x}$
must be separable \cite{gurvits02}. Indeed $\frac{1}{1+D\lambda_{\rho}}\geq\frac{1}{D-1}$,
\textit{i.e.}, ${\lambda}_{\rho}\leq(1-\frac{2}{D})$. Let us also
note that the above result is consistent with the argument presented
in \cite{zy98} that the set of PPT entangled states includes a non-empty
ball.

\textbf{Theorem 2} For every PPT entangled state $\rho$, there is
always a non empty PPT entangled ball of finite radius in its neighbourhood.
Thus almost all PPT entangled states are unconditionally robust.

Denoting the ball $B({\rho}_{x};y_{0}(x))$ in Theorem 1, as $B({\rho}_{x};y_{opt}(x))$
where we have assumed that the entanglement witness considered in
deriving the value of $y_{0}(x)$, is an optimal entanglement witness
$W_{opt}$ and ${\lambda}_{opt}=-~{\textrm{Tr}}\left(W_{opt}\rho\right)$,
consider the following non-empty subset of $({\mathcal{P}}-{\mathcal{S}})$:
\begin{equation}
\mathcal{N}_{PPTBE}={\bigcup}_{\rho\in({\mathcal{P}}-{\mathcal{S}})}{\bigcup}_{x\in]1/(1+D{\lambda}_{opt}),1[}B\left({\rho}_{x};y_{opt}(x)\right).\label{volpptbe}\end{equation}
 It seems that ${\mathcal{N}}_{PPTBE}$ is a proper subset of $({\mathcal{P}}-{\mathcal{S}})$
as it appears that (in particular) the edge states of $({\mathcal{P}}-{\mathcal{S}})$
should not have unconditional robustness properties.

\section{Neighbourhood robustness of bound entangled states from an unextendible
product basis}

In what follows we illustrate all the above-mentioned properties considering
only bound entangled states generated from an unextendible product
basis (UPB) \cite{bennett99}, construct a PPT entangled ball whose
radius can be explicitly found, and use these results to obtain a
lower bound on the volume of PPT entangled states. We further note
that entanglement

\subsection{Bound entangled states from an UPB and entanglement witness \cite{bennett99,terhal00}}

We begin with the definition of bound entangled states constructed
from an UPB.

Let $H$ be a finite dimensional Hilbert space of the form $H_{A}\otimes H_{B}.$
For simplicity we assume that $\dim H_{A}=\dim H_{B}=d.$ Let $S=\left\{ |\omega_{i}\rangle=|\psi_{i}^{A}\rangle\otimes|\varphi_{i}^{B}\rangle\right\} _{i=1}^{n}$
be an UPB with cardinality $\left|S\right|=n$. Let the projector
on $H_{S}$ (the subspace spanned by the UPB), be denoted by $P_{S}={\textstyle \sum\limits _{i=1}^{n}}\left|\omega_{i}\right\rangle \left\langle \omega_{i}\right|.$

\textbf{Lemma 3} \textit{\emph{\cite{bennett99} Let $P_{S}^{\perp}$
be the projector on $H_{S}^{\perp}$ (the subspace orthogonal to $H_{S}$).
Then, the state \begin{equation}
\Omega=\frac{1}{d^{2}-n}\left(I-P_{S}\right)=\frac{P_{S}^{\perp}}{d^{2}-n},,\label{UPB-BE}\end{equation}
 where $D=d^{2},$ is PPT entangled.}}

The state $\Omega$ is the bound entangled state generated from UPB
and will be referred to as the BE-UPB state. In \cite{terhal00},
the following result was proved:

\textbf{Lemma 4} \textit{\emph{Let $S=\left\{ |\omega_{i}\rangle=|\psi_{i}^{A}\rangle\otimes|\varphi_{i}^{B}\rangle\right\} _{i=1}^{n}$
be an UPB. Then \begin{eqnarray}
\lambda & = & \min{\textstyle \sum\limits _{i=1}^{n}}\left\langle \phi_{A}\phi_{B}|\omega_{i}\right\rangle \left\langle \omega_{i}|\phi_{A}\phi_{B}\right\rangle =\min{\textstyle \sum\limits _{i=1}^{n}}\left|\left\langle \phi_{A}|\psi_{i}^{A}\right\rangle \right|^{2}\left|\left\langle \phi_{B}|\varphi_{i}^{B}\right\rangle \right|^{2}\label{lambda}\end{eqnarray}
 over all pure states $\left|\phi_{A}\right\rangle \in H_{A},\left|\phi_{B}\right\rangle \in H_{B}$
exists and is strictly larger than $0.$}}

It was also shown in \cite{terhal00} that in many cases where UPB
states have considerable symmetry, $\lambda$ can be explicitly calculated.

One can accordingly define the entanglement witness operator unnormalized)
that detects UPB-BE states:\begin{equation}
W=P_{S}-\lambda I\label{witness1}\end{equation}
 First of all note that the operator is Hermitian. Next for any product
state $\left|\phi_{A},\phi_{B}\right\rangle \in H,$ $\left\langle \phi_{A},\phi_{B}\right|W\left|\phi_{A},\phi_{B}\right\rangle \geq0$
where the equality is achieved by the product state for which $\left\langle \phi_{A},\phi_{B}\right|P_{S}\left|\phi_{A},\phi_{B}\right\rangle =\lambda$
and from Lemma 4 we know such a product state exists. So, for any
convex combination of projectors on these later product states (let
${\sigma}_{\Omega}$ be one such convex combination), we have ${\textrm{Tr}}\left(W{\sigma}_{\Omega}\right)=0$
and for all separable states $\sigma$, ${\textrm{Tr}}(W\sigma)\geq0.$
One can trivially check that ${\textrm{Tr}}(W\Omega)=-\lambda<0$.
Note that ${\textrm{Tr}}(W)=n-{\lambda}d^{2}$, and hence, we must
have $\lambda<n/{d^{2}}$.

\subsection{PPT entangled balls whose radii can be explicitly calculated and
a lower bound on the volume of PPT entangled states}

From now on, we shall consider the normalized entanglement witness
\begin{equation}
W_{\Omega}=\frac{W}{n-{\lambda}d^{2}}.\label{normwitness}\end{equation}
 The witness operator $W_{\Omega}$ can also detect a large class
of other bound entangled states constructed from UPBs and in particular
the bound entangled states that satisfy the range criterion besides
having less than full rank \cite{bandyopadhyay05}.

Notation-wise, \begin{eqnarray}
{\lambda}_{\Omega} & \equiv & -~{\textrm{Tr}}\left(W_{\Omega}\Omega\right)={\lambda}/(n-{\lambda}d^{2})\label{eq:}\\
p(W_{\Omega}) & = & n\label{eq:}\\
\frac{1}{1+D\lambda_{\rho}} & = & 1-\frac{\lambda d^{2}}{n}\label{eq:}\\
\textrm{Tr}(W_{\Omega}^{+}) & = & \frac{n(1-\lambda)}{n-\lambda d^{2}}\label{eq:}\end{eqnarray}
 Thus all the states,\begin{equation}
{\Omega}_{x}=x\Omega+(1-x)(I/d^{2})\in{\mathcal{F}}_{\Omega}^{(1-{\lambda}d^{2}/n)},\label{eq:}\end{equation}
 (see $\textrm{eqn.\,}$(\ref{frhod})) are PPT entangled for $x\in\left(1-\frac{{\lambda}d^{2}}{n},1\right]$.
Now for the following family of states (see $\textrm{eqn.\,}$(\ref{grhod}))\begin{equation}
\tau(\Omega,\sigma,x,y)\equiv y\sigma+(1-y){\Omega}_{x},\label{eq:}\end{equation}
 we have (using $\textrm{eqn.\,}$(\ref{eqn3})) \begin{equation}
y_{0}(x)=~{\textrm{min}}\left\{ \frac{1-x}{d^{2}-1-x},1-\frac{(1-\lambda)d^{2}}{nx+d^{2}-n}\right\} ,~{\textrm{where}}~x\in\left(1-\frac{{\lambda}d^{2}}{n},1\right].\label{nexteq4}\end{equation}
 Thus we see that for each PPT-BE state ${\Omega}_{x}\in{\tilde{\mathcal{F}}}_{\Omega}^{(1-{\lambda}d^{2}/n)}$
(see $\textrm{eqn.}\,$(\ref{deleted})), there exists a ball $B({\Omega}_{x};y_{0}(x))$
that contains only PPT-BE states, where \begin{equation}
y_{0}(x)=\left\{ \begin{array}{lcl}
\frac{1-x}{d^{2}-1-x} & {\textrm{for}}~{\textrm{all}} & x\in\left]1-\frac{{\lambda}d^{2}}{n},x_{0}\right[,\\
\frac{nx-n+{\lambda}d^{2}}{nx-n+d^{2}} & {\textrm{for}}~{\textrm{all}} & x\in\left[x_{0},1\right[,\end{array}\right.\label{eqn4}\end{equation}
 where\begin{equation}
x_{0}=\frac{n(d^{2}-2)+d^{2}\{1-\lambda(d^{2}-1)\}}{n(d^{2}-2)d^{2}(1-\lambda)}.\label{eq:}\end{equation}
 Thus $y_{0}(x)$ in eqn. (\ref{eqn4}) can be explicitly calculated
for those cases of UPB-BE states $\Omega$ where $\lambda$ can be
explicitly obtained \cite{terhal00}. We therefore have the following
result:

\textbf{Theorem 3} \textit{\emph{For any PPT-BE state $\Omega$ corresponding
to the UPB $S=\left\{ |\omega_{i}\rangle=|\psi_{i}^{A}\rangle\otimes|\varphi_{i}^{B}\rangle\right\} _{i=1}^{n}$
in $d\otimes d$, the PPT-BE states\begin{equation}
{\Omega}_{x}=x\Omega+(1-x)(I/d^{2})\label{eq:}\end{equation}
 where $1-\frac{{\lambda}d^{2}}{n}<x<1$, are unconditionally robust.}}

Given any non-empty subset ${\mathcal{T}}$ of ${\mathcal{D}}$, the
\textit{volume} $|{\mathcal{T}}|$ of ${\mathcal{T}}$ is defined
as the probability of randomly selecting an element of ${\mathcal{D}}$
from ${\mathcal{T}}$. From theorem 5, one can have the following
result regarding lower bounds on the volume of PPT-BE states):

\textbf{Corollary 2} \textit{\emph{$|({\mathcal{P}}-{\mathcal{S}})|\geq|{\mathcal{N}}_{PPTBE}|\geq~{\textrm{max}}\{|B({\Omega}_{x};y_{0}(x))|:1-\frac{{\lambda}d^{2}}{n}<x\leq1\}$,
where $y_{0}(x)$ is given in}} \textit{$\textrm{eqn.\,}$}\textit{\emph{(\ref{eqn4})
and ${\mathcal{N}}_{PPTBE}$ is given in}} \textit{$\textrm{eqn.\,}$}\textit{\emph{(\ref{volpptbe}).}}

\textbf{Remark 3} As a special case of Theorem 2, for every $x\in\left]1-\frac{{\lambda}d^{2}}{n},1\right]$,
the PPT-BE state ${\Omega}_{x}=x\Omega+(1-x)(I/d^{2})$ is maximally
robust. In fact, in this scenario, the corresponding separable state
${\sigma}_{\Omega}$ is taken as any convex combination of all the
product states $|\chi\rangle$ such that $\langle\chi|P_{S}|\chi\rangle=\lambda$.

\textbf{Remark 4} As a special case of Theorem 1, the BE-UPB state
$\Omega$ is robust with respect to ${\mathcal{S}}$. Since every
BE-UPB state is an edge state, this is simply converse of the fact
\cite{lewenstein2001} that every PPT-BE state can be expressed as
a mixture of a separable state with an edge PPT BE state.

\textbf{Theorem 4} \textit{\emph{For every BE-UPB state, there is
an adjacent PPT-BE ball of finite radius, obtained by mixing the BE-UPB
state with all possible separable states.}}

\textbf{Proof:} We focus our attention on the class of states obtained
by mixing an UPB-BE state $\Omega$ with \emph{any separable state}
$\sigma$, \begin{equation}
{\sigma}_{z,\Omega}=z\sigma+\left(1-z\right)\Omega.\label{ourbe}\end{equation}
 The state in $\textrm{eqn.\,}$(\ref{ourbe}) is PPT by construction,
and is inseparable in the domain $z\in[0,\lambda[$ because ${\lambda}_{\Omega}p\left(W_{\Omega}\right)/\left(~{\textrm{Tr}}\left(W_{\Omega}^{+}\right)+{\lambda}_{\Omega}p\left(W_{\Omega}\right)\right)=\lambda$~
$\square$.

\textbf{Remark 6} Robustness of the BE-UPB state $\Omega$, that appears
in Theorem 6, can also be extended with respect to the set ${\mathcal{S}}_{\Omega}$
of all elements $\sigma$ of ${\mathcal{P}}$, where, ${\textrm{Tr}}\left(W_{\Omega}\sigma\right)\geq0$.
Therefore, the state $z\sigma+(1-z)\Omega$ is a PPT-BE state for
all $z\in\left[0,\frac{{\lambda}_{\Omega}}{{\lambda}_{\Omega}+z_{1}}\right[$
where $z_{1}=~{\textrm{inf}}\left\{ {\textrm{Tr}}\left(W_{\Omega}\sigma\right):\sigma\in{\mathcal{S}}_{\Omega}\right\} $.
Or\'{u}s and Tarrach \cite{orus04} have recently shown that for
sufficiently small perturbation of any BE-UPB state $\Omega$ in $d_{1}\otimes d_{2}$
by a density matrix $\sigma,$ ${\sigma}^{T_{B}}>0$ on the subspace
spanned by the kernel of ${\Omega}^{T_{B}}$, the resulting state
is PPT.

\textbf{Remark 7} Numerical methods have already been implemented
to obtain entanglement witnesses for other classes of PPT entangled
states \cite{Brandao2005,BV2004,Eisert2004}. It is quite possible
those witnesses, and the pertinent class of bound entangled states
may be used to obtain lower bounds on the volume of the PPT entangled
class, and a comparsion with our result would be worth studying. However
this is beyond the scope of this work and will be taken up in future.

\section{Multipartite Generalization }

It is easy to generalize the above results to the case of multi-partite
entangled states that are PPT across every bipartition \cite{note4}.
One may consider the set ${\mathcal{P}}_{n}$ corresponding to all
states $\rho$ of an $n$-partite system in the Hilbert space $d_{1}\otimes d_{2}\otimes\ldots\otimes d_{n}$,
where $\rho$ is PPT across every bipartition. Let ${\mathcal{S}}_{n}$
be the subset of ${\mathcal{P}}_{n}$ where each element of ${\mathcal{S}}_{n}$
is fully separable. Thus every $\rho\in\left({\mathcal{P}}_{n}-{\mathcal{S}}_{n}\right)$
has genuine $m$-partite entanglement, where $2\leq m\leq n$. The
set ${\mathcal{S}}_{n}$ is convex and compact (with respect to some
suitable metric). Applying Hahn-Banach theorem, for each $\rho\in\left({\mathcal{P}}_{n}-{\mathcal{S}}_{n}\right)$,
one can obtain a Hermitian operator $W_{\rho}$ (acting on $d_{1}\otimes d_{2}\otimes\ldots\otimes d_{n}$)
such that

(i) ${\textrm{Tr}}\left(W_{\rho}\sigma\right)\geq0$ for all $\sigma\in{\mathcal{S}}_{n}$,

(ii) ${\textrm{Tr}}\left(W_{\rho}\rho\right)<0$,

(iii) ${\textrm{Tr}}\left(W_{\rho}\right)=1$, and

(iv) there exists at least one element ${\sigma}_{\rho}\in{\mathcal{S}}_{n}$
where ${\textrm{Tr}}\left(W_{\rho}{\sigma}_{\rho}\right)=0$.

Thus a result analogous to Theorem 3 holds because there exists a
separable ball $B(I/(d_{1}d_{2}\ldots d_{n});\lambda)$ of finite
radius $\lambda>0$, centred around the maximally mixed state $I/(d_{1}d_{2}\ldots d_{n})$\cite{gurvits03}.
The maximal robustness of $\rho\in\left({\mathcal{P}}_{n}-{\mathcal{S}}_{n}\right)$,
in the direction of ${\sigma}_{\rho}\in{\mathcal{S}}_{n}$ can then
be proved in a straightforward manner. Similarly, robustness of $\rho$
with respect to ${\mathcal{S}}_{n}$ can also be proved analogous
to Theorem 1. The results similar to Lemma 4, Theorem 5, Corollary
2, and Theorem 6 also hold because all completely product pure states
in ${\mathcal{S}}_{n}$ form a compact set. In this case the quantity
${\textrm{inf}}\{\langle\phi|P_{S}|\phi\rangle:|\phi\rangle\langle\phi|\in{\mathcal{S}}_{n}\}$,
where $P_{S}$ is the projector on the subspace spanned by the UPB
$S$, is positive and is attained for some pure state in ${\mathcal{S}}_{n}$.

\vspace{0.5cm} \textbf{Acknowledgements :} Many thanks to Anne Broadbent
and Andrew Scott for careful reading of the manuscipt and helpful
suggestions. SB was supported by iCORE, MITACS, CIFAR and General
Dynamics Canada. Part of this work was completed while SG was visiting
the Institute for Quantum Information Science at the University of
Calgary. Research of SG was funded in part by EPSRC grant GR/87406.
VR was sponsored in part by the Defense Advanced Research Projects
Agency (DARPA) project MDA972-99-1-0017, and in part by the U. S.
Army Research Office/DARPA under contract/grant number DAAD 19-00-1-0172.


\begin{thebibliography}{20}
\bibitem{note2}The state ${\sigma}_{\rho}\in{\mathcal{S}}$ is `nearest' to the state
$\rho\in({\mathcal{P}}-{\mathcal{B}})$ in the sense that the distance
$d(\rho,\sigma)$ of $\rho$ from any element $\sigma$ of ${\mathcal{S}}$
achieves its minimum at the point ${\sigma}_{\rho}$.
\bibitem{note3}Note that this is not true in general for an arbitrary element of
${\mathcal{S}}$. For example, if $|\Psi\rangle\langle\Psi|$ is an
entangled state in ${\mathcal{D}}$, then the (unnormalized) state
$\rho=|\psi\rangle_{A}\langle\psi|\otimes|{\psi}^{*}\rangle_{B}\langle{\psi}^{*}|+\eta|\Psi\rangle\langle\Psi|$
is always entangled for all $\eta>0$ (where we have assumed that
$d_{1}\leq d_{2}$, and for $|\psi\rangle_{A}=\sum_{i=1}^{d_{1}}{\lambda}_{i}|i\rangle_{A}$,
we have $|{\psi}^{*}\rangle_{B}=\sum_{i=1}^{d_{1}}{\lambda}_{i}^{*}|i\rangle_{B}$).
This is because ${\rho}^{T_{B}}$ (the partial transposition of $\rho$
with respect to the subsystem $B$) is not positive semi-definite
when $\eta>0$. Hence there doesn't exist $\lambda>0$ such that the
ball $B(|\psi\rangle_{A}\langle\psi|\otimes|{\psi}^{*}\rangle_{B}\langle{\psi}^{*}|;\lambda)$
contains only separable states.
\end{thebibliography}
\end{document}